\documentclass[WCMS,twocolumn]{USG} 
\usepackage{anyfontsize} %

\usepackage{colortbl}   
\usepackage{xcolor}     

\usepackage{hyperref}
\usepackage{amsmath}
\usepackage{amsfonts}
\usepackage{graphicx}
\usepackage{dcolumn}
\usepackage{braket}
\usepackage{enumitem}
\usepackage[skip=2pt,font=footnotesize]{caption}
\usepackage{accents}
\usepackage{xcolor}
\usepackage{enumitem}

\newcommand{\vecr}{{\mathbf{x}}}
\newcommand{\vecx}{{\mathbf{y}}}

\newcommand{\scriptI}{{\mathcal{I}}}
\newcommand{\scriptF}{{\mathcal{F}}}
\newcommand{\via}{{\textit{via}~}}
\newcommand{\ie}{{\textit{i.e.}~}}
\newcommand{\rmi}[1][]{\mathrm{i}}

\newcommand{\tL}{t_{L}}
\newcommand{\rpL}{r'_{L}}

\newcommand{\rR}{r_{R}}

\newcommand{\tpR}{t'_{R}}

\newcommand{\tC}{t_{c}}

\newcommand{\tpC}{t'_{c}}

\newcommand{\Phip}[1]{\varphi_{#1}^{+}}
\newcommand{\Phim}[1]{\varphi_{#1}^{-}}

\newcommand{\Gdd}{G_{dd}}

\usepackage{steinmetz}
\graphicspath{{./images/}}

\articletype{RESEARCH ARTICLE}%

\received{XX Month Year}
\revised{XX Month Year}
\accepted{XX Month Year}
\journal{}
\volume{0}
\copyyear{Year}
\startpage{1}
\articledoi{XX.XXXX/0000}

\begin{document}
\title{Fast local field calculation in cascaded coupled systems}
\transtitle{Fast local field calculation in cascaded coupled systems}
\author[1]{Isam BEN SOLTANE}[https://orcid.org/0000-0001-7474-6989]
\author[1]{Romain ASSO}
\author[1]{Matthieu DAVY}[https://orcid.org/0000-0002-8023-6456]

\titlemark{Fast local field calculation in cascaded coupled systems}

\address[1]{Univ Rennes, CNRS, IETR - UMR 6164; F-35000 Rennes, France}

\corres{Isam BEN SOLTANE  (\email{isam.ben-soltane@univ-rennes.fr}) ~|~ Matthieu DAVY (\email{matthieu.davy@univ-rennes1.fr})}




\keywords{Green function | S matrix | Coupling | Electromagnetism | Simulation}


\abstract[ABSTRACT]{
Computing the field within a sub-unit of cascaded systems is a recurring problem in wave physics. It is central to gradient-based inverse design, in situations where only a region of the system is modified while the rest remains fixed. Standard approaches either resimulate the entire cascade at every change, or require a detailed characterization of every surrounding sub-system. Both become impractical when many design iterations are needed. Here, we introduce a hybrid method combining the $S$-matrix approach, which propagates fields through cascaded systems, with the Green's function method, which gives the field generated by a source within a characterized system. In this approach, the sub-systems surrounding the region of interest need only be characterized once, after which the local field can be recomputed rapidly for any new configuration. We validate the method on a disordered lossy waveguide, finding excellent agreement with finite-element simulations. We also demonstrate its efficiency on a disordered multimode cavity, where it achieves a fifty-fold speed-up over full-wave simulation and reduces a wideband focusing calculation from eleven hours to ten minutes. This approach opens the way to fast, iterative field calculations in complex systems coupled to their environment, with direct applications to inverse design and wideband wavefront-shaping.}

\maketitle


\section{Introduction}
Numerical methods such as the Finite-Difference Time-Domain (FDTD) and Finite Element Method (FEM)~\cite{ma_Discrete_2005,semouchkina_FDTD_2005,choi_Transmission_2011,smith_Stochastic_2012,demesy_Scattering_2018,horodynski_Antireflection_2022a,niu_Stochastic_2024} are commonly used to accurately describe the propagation of an incident wave within a system. Based on a mesh of the medium, they can provide scattering operators, such as the $S$ matrix, whose coefficients relate the incoming and outgoing fields at the interfaces of the medium~\cite{haynes_Multipole_2011,grigoriev_Singular_2013,pai_Scattering_2021a,horodynski_Antireflection_2022a,zhumabay_Inverse_2026}, as well as the field distribution inside the medium itself. The latter is related to the medium's Green's function~\cite{saillard_Scattering_1990,cangellarisNovelClosedformGreens2000,chenNewGreensFunction2012,zhangGeneralGreensFunction2017,panGreensFunctionsGeophysics2019,boumrarEquivalenceWaveFunction2020}, which gives the field generated by a local source at the position of a receiver.

However, the direct computation of a system with dimensions much larger than the wavelength can quickly become computationally challenging. This challenge is compounded in the presence of multiple scattering, where the outgoing field results from repeated interactions of the incident field with the scatterers. Complex media of this kind arise in many areas of physics, including biological tissue in optics, urban environments at microwave frequencies, or acoustic media~\cite{mosk_Controlling_2012,carminati_Speckle_2015,rotter_Light_2017,delhougne_Coherent_2021,pai_Scattering_2021a,gigan_Roadmap_2022,vynck_Light_2023}. Nevertheless, such a system can often be decomposed into a cascade of several coupled sub-systems, for which the scattering matrices, and the fields within them, are computed separately. This decomposition is of particular interest for a complex medium coupled to other systems, such as imaging devices in biological optics~\cite{liutkus_Imaging_2014,bertolotti_Imaging_2022,gigan_Roadmap_2022}, or antennas~\cite{delhougne_Experimental_2021,davy_Mean_2021,dong_MetamaterialBased_2012,jackson_LeakyWave_2012,milias_MetamaterialInspired_2021,zheng_LeakyWave_2023}.

A particularly demanding case of this decomposition arises in gradient-based inverse design, where only a sub-part of a larger system (the design region) is subject to optimization, while the rest of the system remains fixed and immutable. This is the case, for instance, of an anti-reflective structure introduced to render an already fabricated, disordered scattering medium perfectly transmitting~\cite{horodynski_Antireflection_2022a}, of a wavefront-shaping correction layer compensating for a fixed section of multimode fiber or biological tissue~\cite{mosk_Controlling_2012,gigan_Roadmap_2022}, or of a matching structure optimized to couple an antenna to an otherwise fixed radio-frequency circuit. In all these situations, the gradient of the objective with respect to the design parameters is most efficiently obtained \via the adjoint method~\cite{lalau-keraly_Adjoint_2013,hughes_Adjoint_2018a,elbechari_Adjoint_2022,guillamon_Insitu_2025a}, which requires the overlap, evaluated over the design region only, of the field generated by the physical excitation (the forward field) with that generated by a fictitious source associated with the objective (the adjoint field). Both fields must therefore be known locally within the design region, and must be recomputed at every iteration as the design evolves. At the same time, the surrounding, unmodified part of the system needs to be characterized only once. In this article, we  precisely target this situation: repeatedly retrieving the local field within one sub-system of a cascade, without  resimulating the sub-systems that remain unchanged.

The reconstruction of the full scattering matrix of a cascade from the individual scattering matrices of its sub-systems is a well-established problem, readily solved using $S$-matrix methods for stratified and multiple-scattering media~\cite{neviere_Electromagnetic_1996,stout_Transfer_2002,hugonin_Hybridization_2008,asadova_Tmatrix_2025}. However, retrieving the field within one specific sub-system of interest is far less straightforward. It generally requires either a full simulation of the entire cascade, repeated at every change of that sub-system, or complete knowledge of the boundary conditions and Green's operators of every other sub-system in the chain. Neither option is compatible with situations where the sub-system of interest is modified repeatedly, or where the surrounding sub-systems are only partially characterized.

Here, we introduce a hybrid approach that combines the simplicity of the $S$-matrix method for cascaded coupled systems with the computational efficiency of the Green's function method in simple systems, in order to calculate the field within a complex medium coupled to other systems. We first demonstrate the accuracy of the proposed approach on a lossy metallic waveguide operating at microwave frequencies, split into three coupled sub-systems. Comparing the field computed within the central section using our hybrid approach to that obtained from a full finite-element simulation of the entire waveguide, we find excellent agreement, with an average relative error of only $0.26\%$. We then illustrate the time-efficiency of this hybrid approach on the electric field within a disordered, multichannel metallic cavity at microwave frequencies, into which a random arrangement of cylindrical metallic scatterers is introduced. This complex central sub-system is coupled at both ends to a set of smaller, single-channel waveguides, which are used both to generate the excitation and to measure the resulting field. The Green's function is obtained \via the Coupled Dipole Approximation (CDA)~\cite{chaumet_Modelling_2017,markel_Extinction_2019,chaumetDiscreteDipoleApproximation2022,zhumabay_Inverse_2026}.
Benchmarked against full-wave simulations over $100$ random scatterer configurations, the hybrid approach achieves on average a fifty-fold reduction in computation time, for a moderate relative error of $5.7\%$ arising primarily from the individual FEM and CDA calculations rather than from the hybrid combination itself. We further exploit this speed-up to perform a wideband phase-conjugation simulation, spatially and temporally focusing the field at a chosen position within the cavity from the field computed at $200$ frequencies across a $3$~GHz bandwidth -- a calculation completed in about $10$ minutes with the hybrid approach, against an estimated $11$ hours with a full-wave simulation.

\section{Cascaded coupled systems}

\subsection{Modeling individual systems}

We first consider an individual system which can interact at a set frequency with its environment through two interfaces. The system can be excited at either ends \via the incoming fields $\varphi_1^+$ and $\varphi_2^-$, to generate the outgoing fields $\varphi_1^-$ and $\varphi_2^+$. These localized fields, which act as sources, are linked by the $S$-matrix $S$ of the system:
\begin{equation}
    \begin{bmatrix}
        \varphi_1^- \\
        \varphi_2^+
    \end{bmatrix} 
    =
    S
    \begin{bmatrix}
        \varphi_1^+ \\
        \varphi_2^-
    \end{bmatrix}
    =
    \begin{bmatrix}
        r & t' \\
        t & r'
    \end{bmatrix}
    \begin{bmatrix}
        \varphi_1^+ \\
        \varphi_2^-
    \end{bmatrix}
\end{equation}
where $r$ and $t$ are respectively the reflection matrix at one interface and the transmission matrix from that interface to the other one. The matrices $r'$ and $t'$ are defined similarly for the other interface.

While the $S$-matrix fully determines the fields at the two interfaces of the system, it carries no information about the field \emph{inside} the system itself. Retrieving this internal field, for instance to characterize the local density of states or to identify the position of the scatterers, requires an additional relation between the sources $\varphi_1^+$, $\varphi_2^-$ and the field $\psi$ at any position within the system. This relation is obtained through the assumption of the existence of a linear Green's operator $G$ linking the sources to the field $\psi$ within the system:
\begin{equation}
    \psi = G 
    \begin{bmatrix}
        \varphi_1^+ \\ 
        \varphi_2^-
    \end{bmatrix} 
\end{equation}
The Green's operator $G$ can be expressed in terms of an excitation field $\psi_0$ and a reference Green's operator $G_0$, both defined in a simpler reference system, \via the Lippmann-Schwinger equation~\cite{lippmann_Variational_1950,chaumet_Modelling_2017}. The reference system is typically the system stripped of its scatterers, \ie free space or a homogeneous, lossless waveguide, in which $\psi_0$ and $G_0$ can be computed analytically or at negligible cost. The scatterers are then reintroduced through a source operator $X$, obtained from their individual scattering properties, so that the total field satisfies
\begin{equation}
    \psi = \psi_0 + G_0 X \psi,
    \label{eq:Lippman}
\end{equation}
where $X$ acting on the field $\psi$ generates the secondary sources radiated by the scatterers, which in turn contribute to the field \via the reference operator $G_0$. This self-consistent equation accounts for all orders of multiple scattering between the scatterers. It follows that
\begin{equation}
    \psi = (I-G_0X)^{-1} \psi_0
\end{equation}
The calculation of the field within the system therefore requires the calculation of the reference field $\psi_0$ in the empty system, as well as the knowledge of the secondary source operator $X$.

\subsection{Hybrid $S$-matrix - Green's operator method}

\begin{figure}[ht]
    \centering
    \includegraphics[width=0.99\columnwidth]{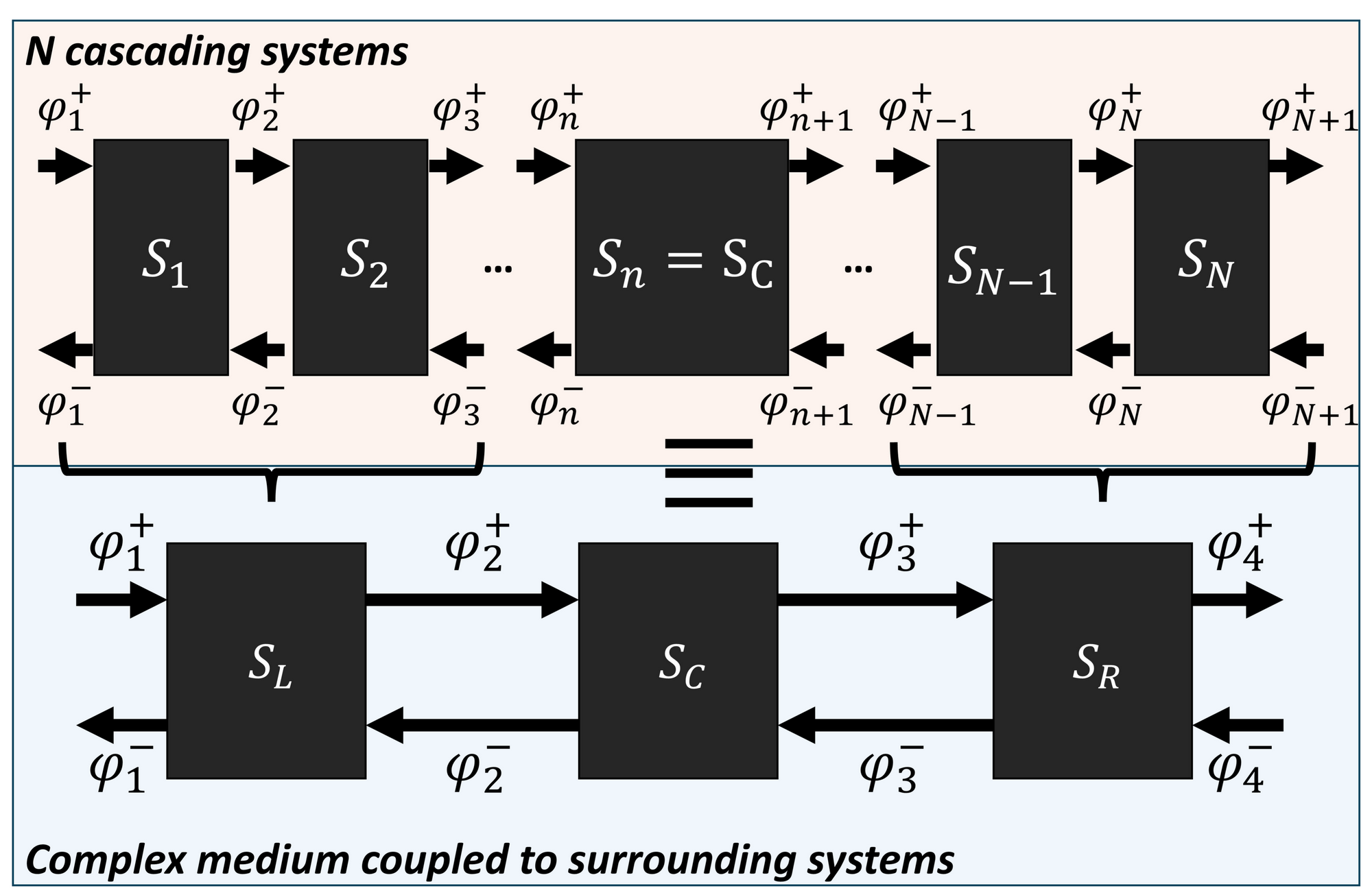}
    \caption{Equivalent representation of $N$ cascaded systems as two systems coupled to a central one. Each system in the cascade has an $S$-matrix $S_N$, some input fields $\Phip{N}$ and $\Phim{N+1}$ and output fields $\Phim{N}$ and $\Phip{N+1}$, which are also shared with the surrounding sub-systems $S_{N-1}$ and $S_{N+1}$. Using the $S$-matrix method, systems $1$ through $n-1$ and $n+1$ through $N$ can be represented as two systems with $S$-matrices $S_L$ and $S_R$, surrounding the $n^{th}$ system with $S$-matrix $S_C=S_n$.}
    \label{fig:fig2}
\end{figure}

We now consider a cascade of $N$ coupled systems, as depicted in Figure~\ref{fig:fig2}. Given the $S$-matrix of every individual sub-system, the input and output fields $\varphi_i^+$, $\varphi_i^-$ at each interface of the cascade can be obtained analytically and efficiently using the $S$-matrix method for stratified and multiple-scattering media~\cite{neviere_Electromagnetic_1996,stout_Transfer_2002,hugonin_Hybridization_2008,asadova_Tmatrix_2025}. 

For two sub-systems characterized by their scattering matrices $S_1$ and $S_2$, the $S$-matrix of the combined system results from the coupling between the two sub-systems, and is given by the Redheffer star product $S = S_1 \star S_2$. This product accounts for all orders of multiple reflection between the two sub-systems, and its coefficients read as
\begin{align}
    r &= r_1 + t'_1 r_2 \left(I - r'_1 r_2\right)^{-1} t_1, \\
    t &= t_2 \left(I - r'_1 r_2\right)^{-1} t_1, \\
    r' &= r'_2 + t_2 r'_1 \left(I - r_2 r'_1\right)^{-1} t'_2, \\
    t' &= t'_1 \left(I - r_2 r'_1\right)^{-1} t'_2.
    \label{eq:redheffer}
\end{align}
Iterating this product over the $N$ sub-systems of the cascade thus yields the $S$-matrix, and hence the fields $\varphi_i^+$, $\varphi_i^-$, at every interface of the full system.

Retrieving the field \emph{inside} one specific complex sub-system $S_n$, however, is far more demanding. Classical methods such as the recursive Green's function approach~\cite{thouless_Conductivity_1981,drouvelis_Parallel_2006} require an effort comparable to that of computing the field over the entire cascade, and rely on complete knowledge of every sub-system, \ie all boundary conditions and Green's operators. In practice, the field in only one sub-system---the $n^{th}$ one---can be typically required, for instance when only this sub-system is subject to non-negligible changes for inverse design approaches while the others remain fixed. Our goal is to compute the field within this sub-system rapidly, without resimulating the entire cascade every time $S_n$ changes. To this end, we combine the $S$-matrix method, applied to the surrounding sub-systems, with a single Green's operator calculation restricted to the $n^{th}$ layer.

We first combine the other $N-1$ sub-systems into two effective systems, with $S$-matrices $S_L$ and $S_R$, surrounding the central complex sub-system described by $S_n \equiv S_C$, as illustrated in Figure~\ref{fig:fig2}. $S_L$ and $S_R$ are obtained by repeated application of the Redheffer star product in Equation~\eqref{eq:redheffer}, to the sub-systems they comprise. Since $S_L$ and $S_R$ do not vary in time, they need only be computed once and the field within these systems does not have to be computed explicitly. Using the same notation as in the previous section, where the $S$-matrix of each system relates its input and output fields, we obtain
\begin{equation}
    \begin{bmatrix} 
        \Phim{k} \\ 
        \Phip{k+1} 
    \end{bmatrix}
    =
    \begin{bmatrix} 
        r_i & t'_i \\ 
        t_i & r'_i 
    \end{bmatrix}
    \begin{bmatrix} 
        \Phip{k} \\ 
        \Phim{k+1} 
    \end{bmatrix}
\end{equation}
with $(i,k) \in \{(L,1),(C,2),(R,3)\}$.

The resulting three-system cascade is equivalent to a succession of three imperfect mirrors, or equivalently, two lossy Fabry-Perot cavities. Multiple reflections between these mirrors generate propagation loops within each cavity, which converge toward a steady state. At this equilibrium, the field $\psi_C$ inside the central complex system can be obtained from a single application of the Green's operator $G$ of the isolated $S_C$ system (\ie in the absence of $S_L$ and $S_R$) to a pair of propagation-invariant, effective sources $\varphi_2^+$ and $\varphi_3^-$:
\begin{equation}
    \psi_C = G
    \begin{bmatrix}
        \varphi_2^+ \\ 
        \varphi_3^-
    \end{bmatrix}
    \label{eq:6}
\end{equation}
These effective sources encapsulate all the multiple reflections occurring within the full cascade, and are expressed in terms of the actual sources of the full system, $\varphi_1^+$ and $\varphi_4^-$, as
\begin{equation}
    \Phip{2} = \Gamma_L \left[ \tL \Phip{1} + \rpL \tpC \gamma_R \tpR \Phim{4} \right]
    \label{eq:7}
\end{equation}
and
\begin{equation}
    \Phim{3} = \Gamma_R \left[ \tpR \Phim{4} + \rR \tC \gamma_L \tL \Phip{1} \right],
    \label{eq:8}
\end{equation}
where the coupling matrices $\gamma_L$, $\gamma_R$, $\Gamma_R$ and $\Gamma_L$ are defined as
\begin{gather}
    \gamma_R = (I - r_Rr'_C)^{-1}\\
    \gamma_L = (I - r'_Lr_C)^{-1}
\end{gather}
and 
\begin{gather}
    \Gamma_L = \left[\gamma_L^{-1} - r'_L t'_C \gamma_R r_R t_C\right]^{-1}\\
    \Gamma_R = \left[\gamma_R^{-1} - r_R t_C \gamma_L r'_L t'_C\right]^{-1}
\end{gather}
The coupling matrices are brought about by the changes of interfaces or impedance mismatches between the different systems, introducing different Fabry-Perot cavities. The first coupling matrix $\gamma_R$ is associated with the cavity $\Omega_R$ made of the reflection coefficients $r'_C$ and $r_R$. It describes the round trips of the field in between the central and right sub-systems. The second coupling matrix $\gamma_L$ plays a symmetrical role for the cavity $\Omega_L$ made of the central and left sub-systems, with reflection coefficients $r'_L$ and $r_C$.
The two other coupling matrices are more complex, as they both involve the previous two Fabry-Perot cavities. $\Gamma_L$ corresponds to the round trips of the field in the cavity $\Omega_L$ to which is subtracted the part of the field resulting from the propagation and coupling in $\Omega_R$. Similarly, $\Gamma_R$ describes the round trips in the cavity $\Omega_R$ after removing the contribution from $\Omega_L$. The derivation of Equations~\eqref{eq:7} to ~\eqref{eq:8}, is detailed in the Supporting Information. 

In practice, the field $\psi_C$ is evaluated at a finite set of $D$ positions $\vecx_m$ within the sub-system. Equation~\eqref{eq:6} can then be recast as a Green matrix $g_{XM}$, relating the source inputs $\Phip{2}$ and $\Phim{3}$, expressed in a chosen source basis, to the output field at the $D$ positions:
\begin{equation}
    \begin{bmatrix}
        \psi_C(\vecx_1) \\
        ... \\
        \psi_C(\vecx_D)
    \end{bmatrix} = 
    g_{XM} 
    \begin{bmatrix}
        \Phip{2} \\
        \Phim{3}
    \end{bmatrix}
\label{eq13}
\end{equation}
where the subscripts $X$ and $M$ denote, respectively, the spatial grid and the source basis.

\begin{figure*}[!t]
    \centering
    \includegraphics[width=0.99\textwidth]{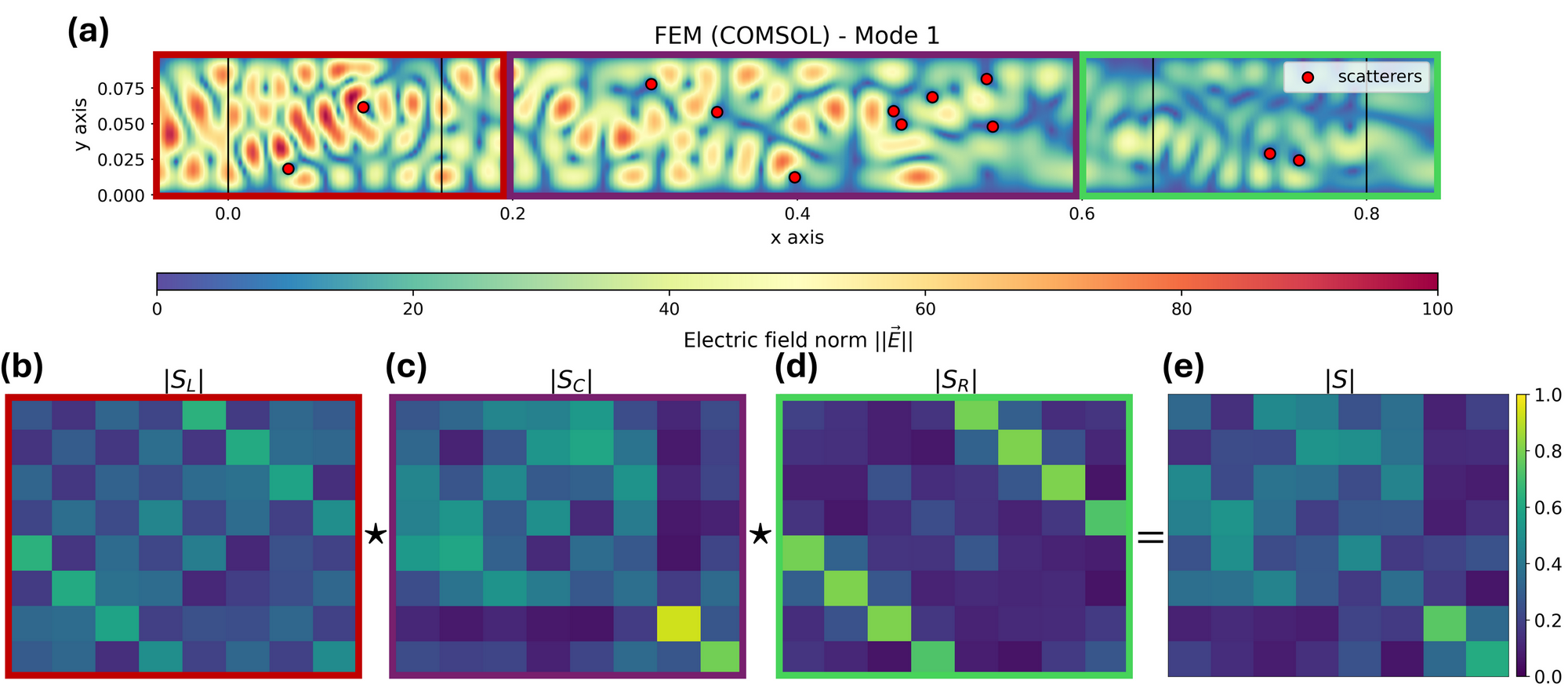}
    \caption{Metallic waveguide operating in the radio frequencies at $7GHz$ (a), split into three parts. The left part is a succession of three layers: a layer of air, an absorbing medium with relative permittivity $\varepsilon=1.5-0.02\rmi$, within which $2$ metallic scatterers are embedded, and another layer of air. It is associated with the $S$-matrix $S_L$ (b). The central part contains $8$ metallic scatterers and has the $S$-matrix $S_C$ (c). The third part is similar to the first one and has the $S$-matrix $S_R$ (d). The total $S$-matrix $S$ is obtained \via the Redheffer product $S=S_L \star S_C \star S_R$ (e).}
    \label{fig:fig3}
\end{figure*}

The hybrid approach proposed here can thus be summarized in three steps:
\begin{enumerate}[label=(\roman*)]
    \item compute the Green matrix $g_{XM}$ of the isolated complex medium, which gives the field generated by any source within that medium;
    \item couple this medium to its surrounding sub-systems using the $S$-matrix method, assuming all $S$-matrices are known;
    \item project the actual sources of the full system onto the chosen basis and propagate them through the coupled system.
\end{enumerate}
Of these three steps, the first is often the most time-consuming, and therefore sets the overall computational cost of the method. Crucially, however, it needs to be performed only once: the efficiency of the approach then comes from repeating steps (ii) and (iii) ---both usually fast--- for any new source or configuration of the surrounding sub-systems, whereas classical approaches require a full simulation for every new excitation.

We demonstrate the ability of the hybrid approach to accurately calculate the field in the test case of a simulated, lossy 2D waveguide operating in the RF.

\section{Error characterization in a waveguide with absorption}

\begin{figure*}[!t]
    \centering
    \includegraphics[width=0.99\textwidth]{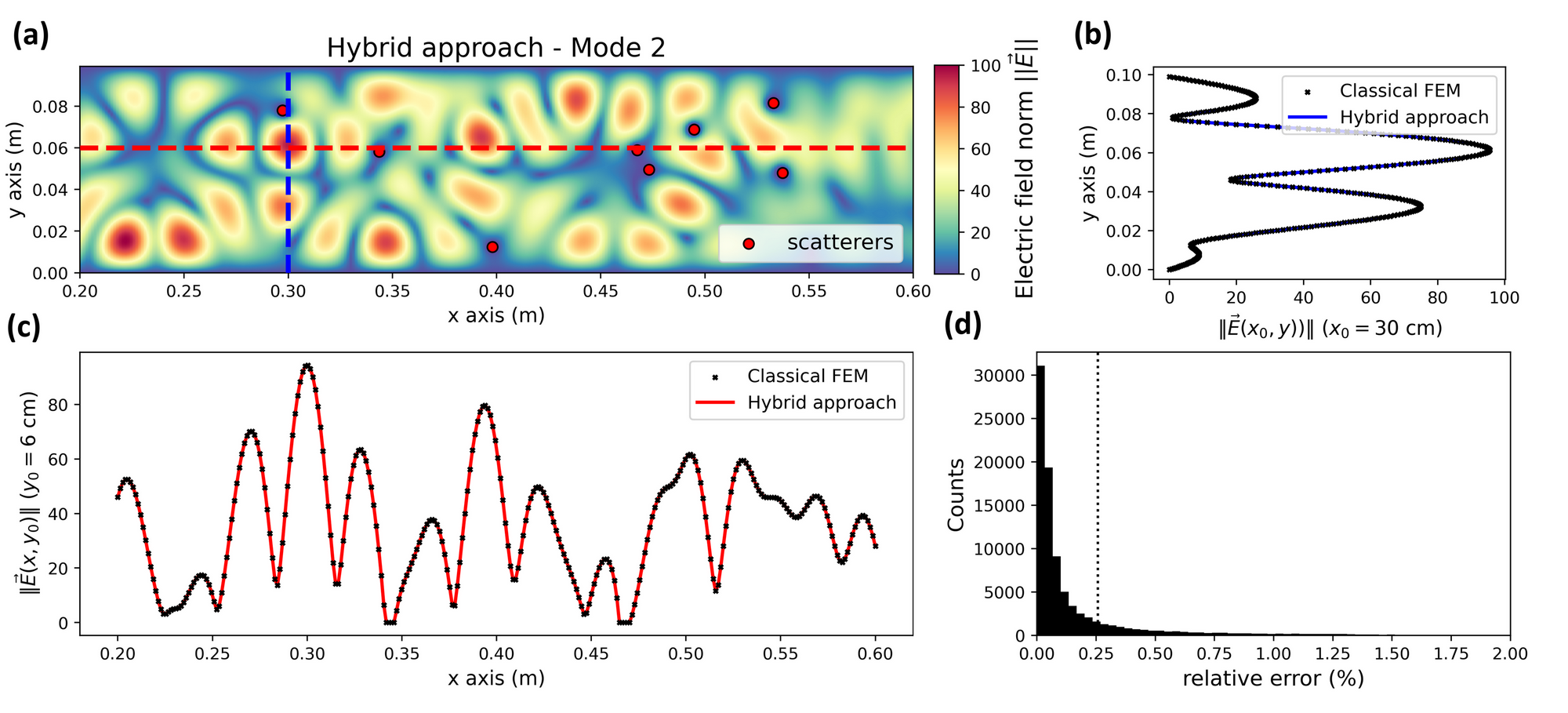}
    \caption{Local field calculation in the central part of the cavity described in Figure~\ref{fig:fig3}. (a) Electric field amplitude map obtained \via the classical full FEM (crosses) and the hybrid approach (full, blue line) for an excitation with the second mode. (b) Electric field curve along the $y$ axis of the waveguide using the full FEM (crosses) and the hybrid approach (full, red line), at the longitudinal position $x_0=30$ cm (in the complete waveguide). (c) Electric field curve along the $x$ axis, at the transverse position $y=6$ cm. (d) Histogram of the relative $L1$ error of the field between the hybrid method and the classical full FEM for the second mode excitation. The curves of the classical FEM and the hybrid approach matches well, and the quantification of the error shows a low relative error average at $0.26\%$ (dotted vertical line).}
    \label{fig:fig4}
\end{figure*}

We consider the 2D electromagnetic waveguide depicted in Figure~\ref{fig:fig3}, supporting $4$ modes at $7$ GHz. It is split into three parts. The first and third parts each contain $2$ metallic scatterers embedded within an absorbing medium of relative permittivity $\varepsilon_a=1.5 - 0.005\rmi$. The medium itself is surrounded by air. The $S$-matrices $S_L$ and $S_R$ of these two parts are depicted in Figure~\ref{fig:fig3} (b) and (d). The central part of the waveguide contains $8$ metallic scatterers embedded within air, and its $S$-matrix $S_C$ is represented in Figure~\ref{fig:fig3} (c). The total $S$-matrix, depicted in Figure~\ref{fig:fig3} (e) is obtained by successively applying the Redheffer product $S=S_L \star S_C \star S_R$.

To evaluate the performance of the method and show that the process does not introduce errors beyond those brought about by the Green function and $S$-matrix methods, we use COMSOL Multiphysics FEM for both the classical full simulation approach and our hybrid approach. We compute the field within the central part of the cavity using the hybrid approach by calculating with the FEM the three $S$-matrices $S_L$, $S_C$ and $S_R$, as well as the Green matrix $g_{XM}$ introduced in Equation~\eqref{eq13}. 

We compare the local field retrieved in this manner to the one obtained through a FEM simulation of the whole waveguide. The results are presented in Figure~\ref{fig:fig4} for an excitation of the whole cavity with the second mode. The full field distribution map obtained using the hybrid approach is shown in Figure~\ref{fig:fig4}(a). The curve of the field amplitude are plotted for two slices at $x=30$~cm in Figure~\ref{fig:fig4} (b) and $y=6$~cm in Figure~\ref{fig:fig4}(c), for both the hybrid and classical approach, qualitatively demonstrating an excellent agreement between the two approaches. The accuracy of the hybrid approach is further quantified through the relative error histogram in Figure~\ref{fig:fig4}(d), with an average error between the two approaches of $0.26 \%$. Overall, the relative error histogram exhibits a concentration of the error distribution close to $0\%$, and is evidence of the accuracy of the hybrid approach. Additional results are presented in the Supporting Information for other modes.

The three step process provided for the hybrid Green's function and $S$-matrix method approach thus allows for an accurate estimation of the field locally, in a part of the system. By replacing the calculation of $g_{XM}$ and $S_C$ with quicker ones or simulations, the full potential of the hybrid approach for both accurate and fast computations becomes evident. We demonstrate this efficiency of the hybrid approach in the case of a disordered multimode cavity. We calculate $g_{XM}$ and $S_C$ with the CDA, resulting in fast and accurate computation of the field, albeit with a higher relative error.

\section{Application to a disordered multimode cavity}

\begin{figure*}[!t]
    \centering
    \includegraphics[width=0.99\textwidth]{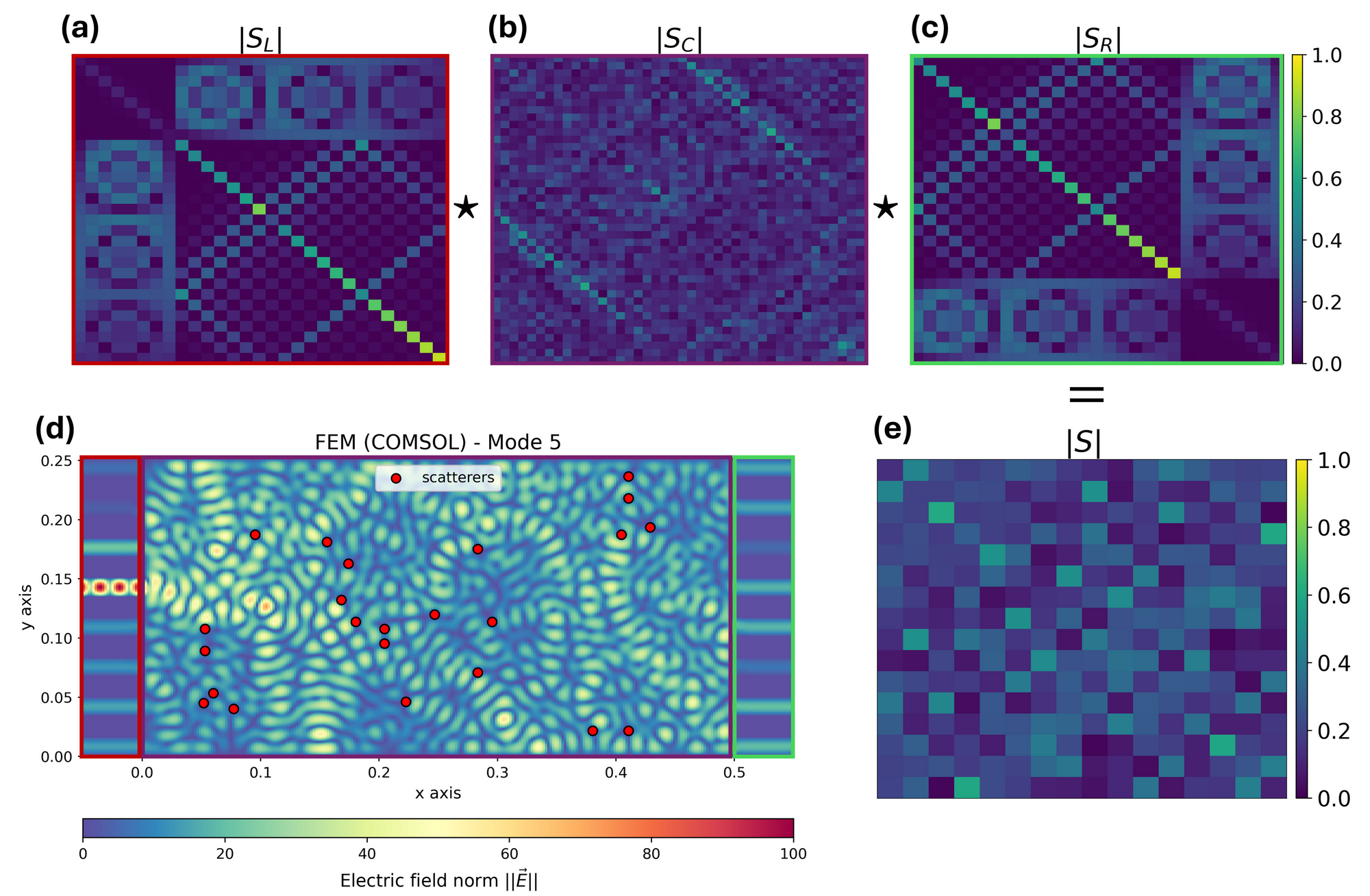}
    \caption{Simulated metallic RF cavity, split into three parts. The central part, where the electric field is to be calculated, is a waveguide of $S$-matrix $S_C$ in which metallic scatterers are randomly distributed. It is surrounded by two symmetrical sub-systems, each composed of $N_w=8$ smaller waveguides coupled to the main one, with $S$-matrices $S_L$ and $S_R$. The cavity has a thickness of approximately $8$ mm, and the full system is modelled as a $2D$ system operating at $13$ GHz.}
    \label{fig:fig5}
\end{figure*}

We apply our hybrid approach to a two-dimensional waveguide coupled to $2N_w = 16$ smaller openings, each a single-channel waveguide, used both to inject the incident field into the main waveguide and to measure the resulting field, as shown in Figure~\ref{fig:fig5}. The central waveguide has a total length $H_{\text{guide}}=50$ cm and a width $W_{\text{guide}}=25.2$ cm, and the full system operates at $13$ GHz in the transverse-electric polarization. A random arrangement of metallic scatterers of diameter $d=6$ mm is placed inside the central waveguide. 

Crucially, the spacing between the single-channel waveguides at the left and right interfaces is metallic, so that waves entering or leaving the cavity through this spacing undergo chaotic propagation~\cite{davy_Selectively_2018,davy_Mean_2021}. The complete system can therefore be viewed as a multichannel cavity with $2N_w$ coupled channels. To model it within our hybrid framework, we decompose the cavity into three sub-systems: the first comprises the $N_w$ single-channel waveguides at the left interface together with the portion of the central waveguide within a distance $l_0$ of that interface; the second is the central waveguide filled with the metallic scatterers; and the third, at the right interface, is the mirror image of the first. These three sub-systems are characterized by their scattering matrices $S_L$, $S_C$, and $S_R$, shown in {Figure~\ref{fig:fig5}} (a), (b) and (c), and the full $S$-matrix $S$ depicted in {Figure~\ref{fig:fig5}} (e) is the Redheffer product of the three subsystems, \ie  $S = S_L\star S_C \star S_R$. 

The left and right $S$-matrices $S_L$ and $S_R$ are obtained \via FEM simulations in COMSOL Multiphysics, while the Green matrix $g_{XM}$ and the central $S$-matrix $S_C$ are computed using the Coupled Dipole Approximation (CDA)~\cite{chaumet_Modelling_2017,markel_Extinction_2019,chaumetDiscreteDipoleApproximation2022,zhumabay_Inverse_2026}, which provides a direct, discrete realization of the reference operator $G_0$ introduced in Equation~\eqref{eq:Lippman}. The CDA yields the analytically known Green's function of the empty waveguide together with the pointwise dipole interactions that play the role of the secondary-source operator $X$, so that the self-consistent Lippmann-Schwinger equation is solved exactly within this discrete representation (see the derivation below for details).

The $N_s$ metallic scatterers within the central waveguide are represented as point dipoles located at positions $\vecr_k$, each characterized by a polarizability $\alpha_k$. Together, they play the role of the secondary-source operator $X$, acting on the total field $\psi$. In empty waveguides, both $\psi_{\mathrm{inc}}$ and the associated reference Green's operator $G_0$ are known analytically as a sum over the guide's propagating and evanescent modes. The total field then reads
\begin{equation}
    \psi(\mathbf{r}) = \psi_{\mathrm{inc}}(\mathbf{r}) + \sum_{k=1}^{N_s} G_0(\mathbf{r},\vecr_k)\, p_k,
    \label{eq:cda_total_field}
\end{equation}
where $p_k=\alpha_k \psi(\vecr_k)$ is the dipole moment induced at $\vecr_k$, and the sum over dipoles plays the role of the term $G_0 X \psi$ in Equation~\eqref{eq:Lippman}. 

The vector of dipole moments $\mathbf{p}_m$ induced by mode $m$ is obtained by solving the self-consistency condition analogous to $\psi = (I-G_0X)^{-1}\psi_0$, here written as a linear system
\begin{equation}
    \left(\mathrm{diag}(\alpha^{-1}) - \Gdd\right)\mathbf{p}_m = \mathbf{G}_{0m},
    \label{eq:cda_W}
\end{equation}
where $\Gdd$ is the matrix of pairwise Green's functions between dipoles, $[\Gdd]_{ij}=G_0(\vecr_j,\vecr_i)$.

Rather than working with the field in real space, we directly evaluate the scattering matrix in the basis of the waveguide's flux-normalized propagating modes. For an incoming mode $m$ and an outgoing mode $n$, the corresponding element of the empty-waveguide-corrected $S$-matrix is
\begin{equation}
    S_{nm} = S_{0,nm} + \mathbf{G}_{0n}^{T}\mathbf{p}_m,
    \label{eq:cda_Snm}
\end{equation}
where $S_0$ is the $S$-matrix of the empty waveguide and $\mathbf{G}_{0n}$ is the vector of empty-waveguide field values at the dipole positions, generated by the $n^{\text{th}}$ incoming mode.  Combining Equations~\eqref{eq:cda_W} and~\eqref{eq:cda_Snm} over all mode pairs yields the compact expression
\begin{equation}
    S = S_0 + G_0^T \left(\mathrm{diag}(\alpha^{-1}) - G_\mathrm{dd}\right)^{-1} G_0,
    \label{eq:cda_S_full}
\end{equation}
with $G_0$ the $N_s \times 2N$ matrix collecting the vectors $\mathbf{G}_{0n}$ for all $2N$ waveguide modes. 

The Green matrix $g_{XM}$ follows from the same dipole moments, projected instead onto the $D$ spatial positions $\vecx_m$, where the field is to be evaluated, using the same reference Green's function $G_0(\mathbf{r},\mathbf{r}')$ but with $\mathbf{r}$ taken on the output grid rather than on an outgoing waveguide mode: $g_{XM}$ and $S_C$ are thus two different projections of the same underlying dipole solution, explicitly given by
\begin{equation}
    \psi(\vecr) = \psi_{inc}(\vecr) + G_r^T \left(\mathrm{diag}(\alpha^{-1}) - G_\mathrm{dd}\right)^{-1} G_0,
    \label{eq:cda_gMX}
\end{equation}
This gives $g_{XM} = G_r^T \left(\mathrm{diag}(\alpha^{-1}) - G_\mathrm{dd}\right)^{-1} G_0$, so that $g_{XM}$ and $S_C$ (Equation~\eqref{eq:cda_S_full}) share the same resolvent $\left(\mathrm{diag}(\alpha^{-1}) - G_\mathrm{dd}\right)^{-1}$ and differ only through the vector $G_0$ (projection onto waveguide modes) being replaced by $G_r$ (projection onto the $D$ output positions), with $[G_r]_{m,k}=G_0(\vecx_m,\vecr_k)$ being the Green function from the dipole position $\vecr_k$ to the grid position $\vecx_m$. The diagonal terms $G_0(\vecr_k,\vecr_k)$ of $G_\mathrm{dd}$, which enter Equation~\eqref{eq:cda_W} through $\mathrm{diag}(\alpha^{-1})$, guarantee the unitarity of $S$~\cite{markel_Extinction_2019}; equivalently, they can be absorbed into a renormalized polarizability $\tilde\alpha_k = \alpha_k/\left(1-\alpha_k G_0(\vecr_k,\vecr_k)\right)$, in which case the diagonal of $G_\mathrm{dd}$, is set to zero. For our lossless scatterers, the bare polarizabilities are purely imaginary, with a sign that depends on the scatterer type (dielectric or metallic). Since the reference Green's function $G_0(\mathbf{r},\mathbf{r}')$ is evaluated as a sum over waveguide modes, the return term $G_0(\vecr_k,\vecr_k)$ diverges logarithmically as the number of modes retained in the sum, $N_{\mathrm{eff}}$, increases~\cite{rivero_GreensFunctionDefect_2021}; we therefore truncate the modal sum at $N_{\mathrm{eff}}=6N$, well beyond the number of propagating modes $N$, so as to capture the dominant evanescent coupling between dipoles while avoiding this divergence. Additionally, because the scattering cross-section of the metallic cylinders used here is large compared to the wavelength, a single dipole per scatterer is insufficient; each cylinder is instead represented by six dipoles distributed along its boundary. The polarizability values used for the metallic and dielectric scatterers are calibrated against full-wave FEM simulations in COMSOL Multiphysics.

This hybrid approach, combining FEM and CDA, is first benchmarked against a classical full-FEM simulation over a range of random scatterer configurations within the cavity, before being applied to a wideband, temporal focusing scenario. Depending on the chosen mesh size, computing $S_L$ and $S_R$ takes between $20$ s and $130$ s; however, this computation is performed only once, and the resulting matrices are then reused for all $N_w$ excitations and all $100$ scatterer configurations considered below.

\setcounter{figure}{5}
\begin{figure*}[!b]
    \centering
    \includegraphics[width=0.99\textwidth]{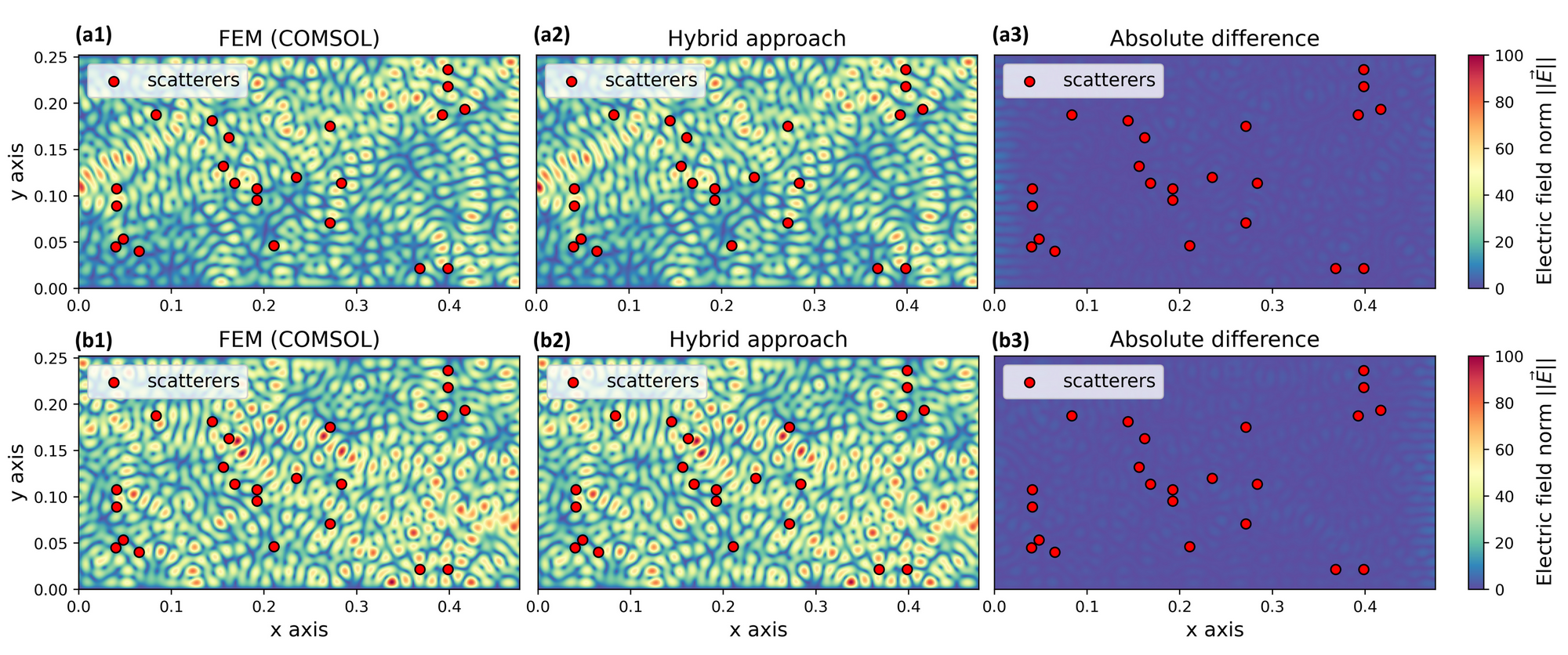}
    \caption{Field distribution within the RF waveguide depicted in Figure~\ref{fig:fig5}, for a random distribution of metallic scatterers (red circles), with an excitation from the $4^{th}$ small waveguide on the left (a1-a3), and the second one on the right (b1-b3). The firs column corresponds to the classical approach using the FEM (a1), (b1), the second one is the hybrid approach (a2), (b2), and the third one is the absolute difference map between the two approaches (a3), (b3). The two distributions are qualitatively in good agreement. The error maps indicate that the dominant sources of errors are near the interface where the excitation occurs, indicating evanescent modes.}
    \label{fig:fig7}
\end{figure*}

\subsection{Benchmark at a fixed frequency}

We calculate the field brought about by the excitation of each of the $N_w$ smaller waveguides, for $100$ random configurations of varying numbers of metallic scatterers. We consider two metrics which are summarized in Figure~\ref{fig:fig6}: the  calculation time for each configuration using the hybrid approach and full FEM simulation in Figure~\ref{fig:fig6} (a), and the relative $L_2$ error of the former with respect to the latter in Figure~\ref{fig:fig6} (b).

\setcounter{figure}{4}
\begin{figure}[!h]
    \centering
    \includegraphics[width=0.95\columnwidth]{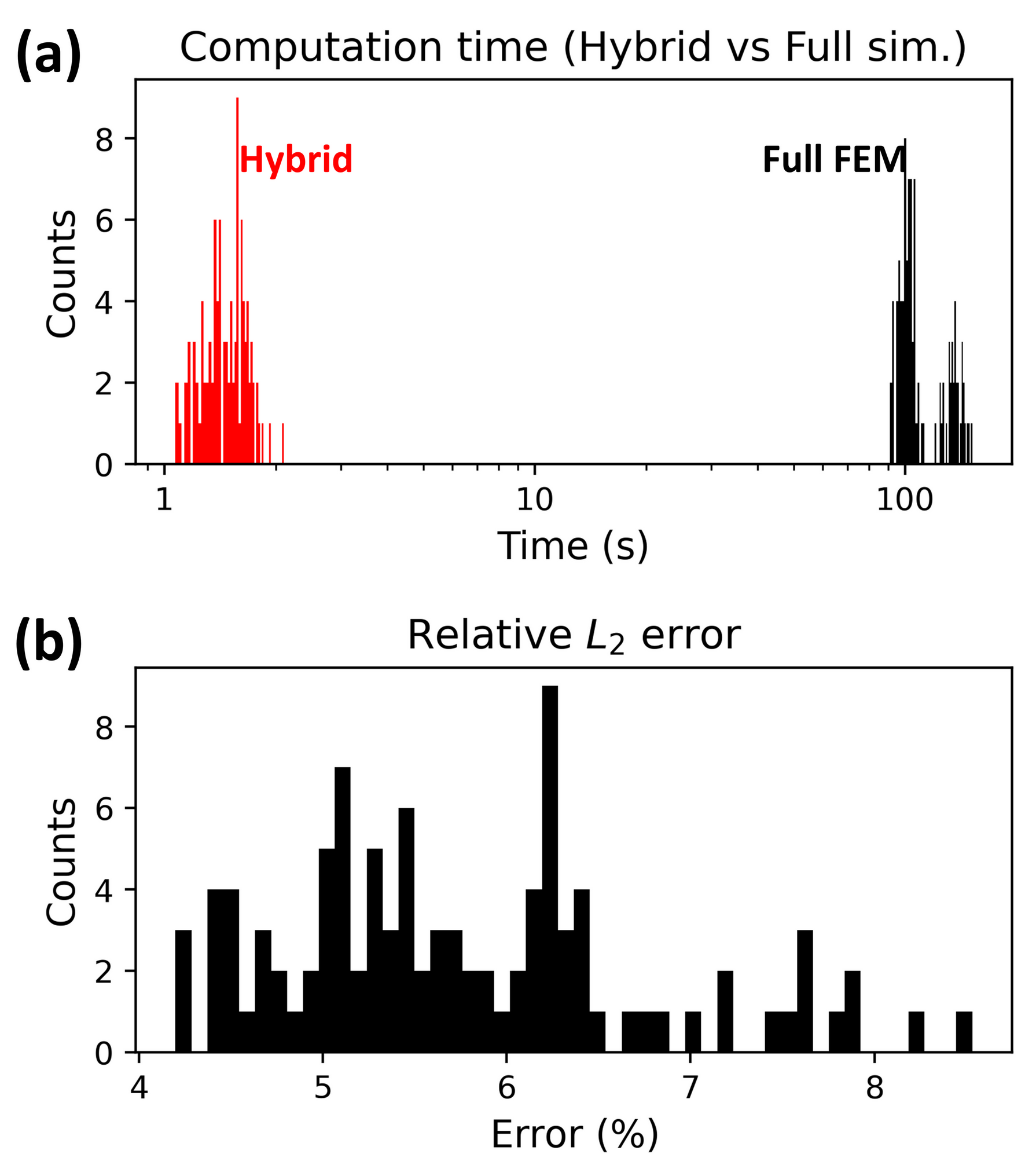}
    \caption{Benchmark of the hybrid approach applied to the electric field in the central part of the cavity introduced in Figure~\ref{fig:fig5}. (a) A total of $100$ random configurations of scatterers in the cavity are tested, and the electric field computation time is calculated and compared to that of full FEM simulations. (b) The relative $L_2$ error with respect to the classical full FEM simulations is also computed. Overall, the hybrid approach provides a good approximation of the field with an average relative error of less than $6\%$, for a fifty-fold improvement of the calculation time.}
    \label{fig:fig6}
\end{figure}

The hybrid approach assumes that the matrices $S_L$ and $S_R$ are already known through simulations or measurements. Consequently, their calculation time is not accounted for in Figure~\ref{fig:fig6}(a). Nonetheless, for the chosen full COMSOL FEM simulation and the CDA, we observe on average a fifty-fold improvement of the computation time for the hybrid approach compared to the classical one. In addition, the full potential of the hybrid method was not used, and the simulations could be further sped up by computing the field for all $2N_w$ excitations of a given scatterer configuration simultaneously, since they share the same $S_L$, $S_R$ and $S_C$ matrices.

The relative errors average at $5.74\%$. Most errors result from the chosen simulation methods rather than from the hybrid approach itself. Two effects dominate: 1) evanescent modes that are not captured by the $S$-matrices computed with the FEM, and 2) inaccuracies intrinsic to the CDA representation of the scatterers. This claim is supported, in the case of an arbitrary configuration, by Figure~\ref{fig:fig7}, where we qualitatively compare the two approaches and show the absolute error over a grid in the waveguide. We observe a higher error close to the interfaces of the waveguide, which globally decreases as we move closer to the centre, consistent with an evanescent-mode effect, while, although less visible, the error also spikes within the scatterers, consistent with the expected accuracy limits of the CDA.

\setcounter{figure}{6}
\begin{figure*}[!b]
    \centering
    \includegraphics[width=0.99\textwidth]{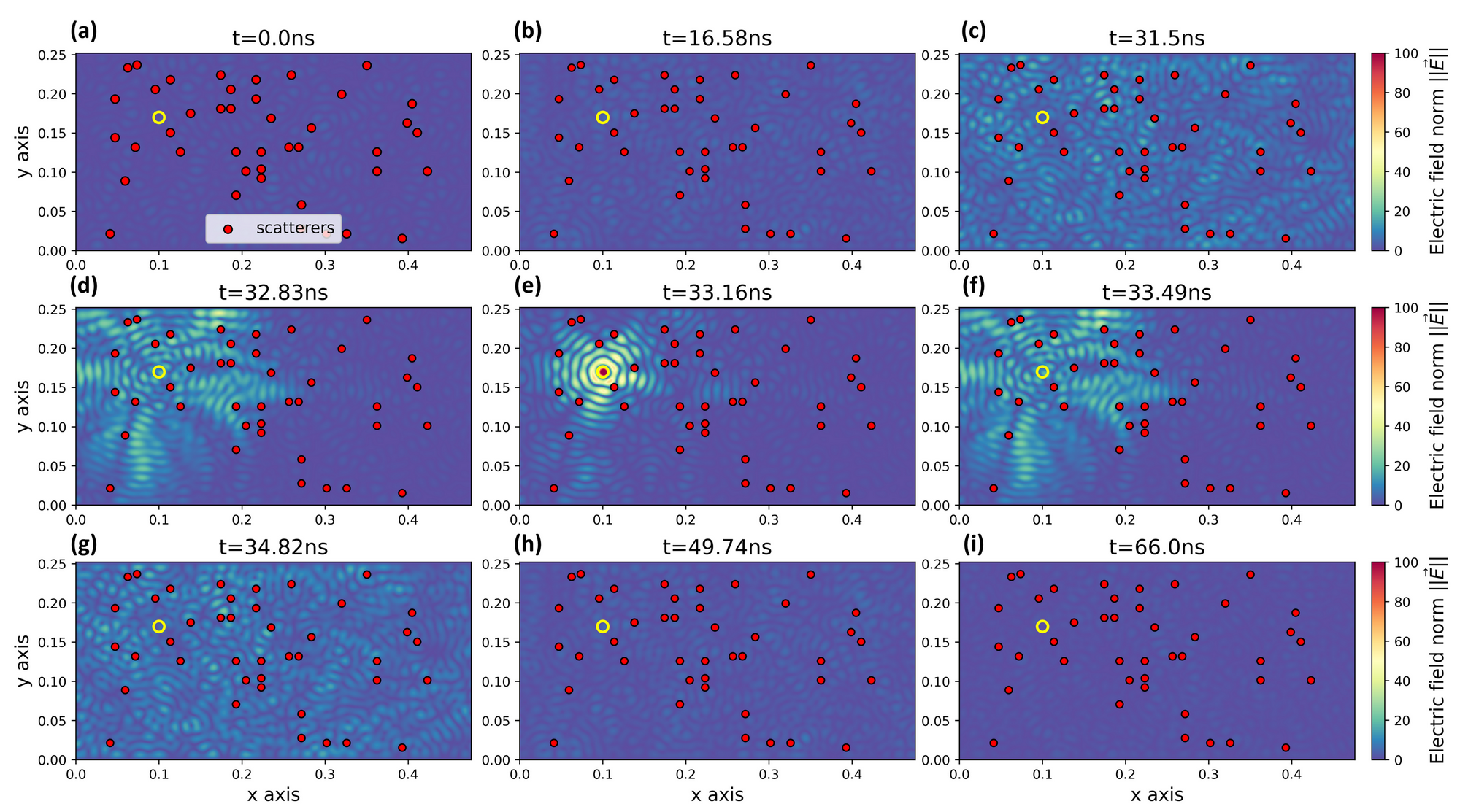}
    \caption{Electric field distribution in the waveguide portion of the RF cavity at different times, for a specific distribution of metallic scatterers. The wideband excitation is obtained \via phase conjugation of the row transmission vector from the small input waveguides to the highlighted position $\vecx_0$ in the waveguide, over the $12.5$ to $13.5$ GHz range. We observe a space and time focusing of the field at the position $\vecx_0$ at $t=33.16$ns, corresponding to the middle of the simulation time range. The field maps are calculated in $10$min using the hybrid approach, while the runtime is estimated to at least $11$ hours with the classical FEM approach.}
    \label{fig:fig8}
\end{figure*}

\subsection{Wideband phase conjugation for intensity maximization in time}

The speed of the hybrid approach allows for applications requiring several simulations, such as wideband calculations. We demonstrate the efficiency of the method in the specific case of field focusing at a specific position within the waveguide.

At a given frequency $\nu$, we consider the $1\times 2N_w$ transmission matrix $t(\nu)$ from each of the $2N_w$ smaller waveguides connected to the cavity to an arbitrarily chosen position $\vecx_0$ of the waveguide. where we want the field intensity to be maximized through time. This rank $1$ matrix is a row vector. Its transpose conjugate is thus its only singular vector (up to to scaling factor), which maximizes the field at $\vecx_0$ for the frequency $\nu_0$. The resulting field $\psi(\nu, \vecx_0)$, obtained through phase conjugation, reads as
\begin{equation}
    \psi(\nu, \vecx_0) = t(\nu) t^H(\nu).
\end{equation}
By calculating $\psi(\nu,\vecx_0)$ similarly at every frequency $\nu$ within a spectral range $\Omega=\left[\nu_1, \nu_2\right]$, we are thus maximizing the quantity
\begin{equation}
    \scriptI = \int_{\nu_1}^{\nu_2} \psi(\nu,\vecx_0) d\nu = \sum_{k=1}^{N_w} \int_{\nu_1}^{\nu_2} |t_k(\nu)|^2 d\nu
    \label{eq:11}
\end{equation}
with $t_k$ the component of $t$ corresponding to the excitation of the $k^{th}$ small waveguide. For $\Omega$ large enough, or if $t$ is close to $0$ outside of $\Omega$, the Plancherel theorem~\cite{plancherel_Contribution_1910} allows us to recast Equation~\eqref{eq:11} into the temporal domain as
\begin{equation}
    \scriptI = \sum_{k=1}^{N_w} \int_{-\infty}^{\infty} |\scriptF^{-1}[t_k](t)|^2 dt = \scriptF^{-1}[\psi](0,\vecx_0)
\end{equation}
with $\scriptF$ the inverse Fourier transform operator. We are thus effectively maximizing the transmission from the $N_w$ waveguides to the position $\vecx_0$ in the temporal domain, by maximizing the field amplitude at the time origin.

We choose $\Omega$ ranging from $12$ GHz to $15$ GHz, with $200$ frequencies uniformly distributed over that range. The hybrid approach is applied at each frequency to retrieve the field and apply the phase conjugation method. The full process runs in approximately $10$ min, while its runtime is estimated to at least $11$ hours with a full COMSOL FEM approach. The result is presented in Figure~\ref{fig:fig8} at nine different times.

The field focuses at the circled position $\vecx_0$ at $t=33.16$ ns, roughly corresponding to the middle of the temporal window obtained \via discrete inverse Fourier transform of the field calculated over $\Omega$. The ratio of the amplitude at $\vecx_0$ over the average amplitude in the waveguide, at $t=32.16$ ns, is of $42$. This complements the quantitative distribution of the amplitude presented in Figure~\ref{fig:fig8} in highlighting the success of the temporal focusing at $\vecx_0$. 

The full temporal simulation can be found in the Zenodo repository \cite{bensoltane_zenodo_2026}.

\section{Conclusion}

The calculation of the field within a sub-system of cascaded systems with classical numerical approaches rapidly becomes time-consuming. In particular, studies involving several modifications of that system require a full simulation at every step. In this paper, we have introduced a hybrid approach leveraging the efficiency of the $S$-matrix method for stratified media, combined with the Green function approach to calculate the field generated by a source, to drastically reduce the time needed to compute the field within a complex media coupled to other surrounding systems. This approach depicts the full system as a succession of two Fabry-Perot cavities for which analytical expressions of the coupling are used to rapidly calculate the sources exciting the sub-system of interest. The Green function method then allows for a propagation of these effective sources to any position of that sub-system. We have shown the accuracy of this hybrid approach by comparing it to a full finite-element-method simulation in a lossy waveguide, where an average relative error of $0.26\%$ was obtained. We have then demonstrated the time-efficiency of the method in several metallic cavity containing random distribution of metallic scatterers. We have shown a fifty-fold improvement of the computation time on average with respect to full FEM simulations.  In the specific case of the wideband phase conjugation approach then considered, we have provided further evidence of the potential of the approach by reducing a computation time estimated to $11$ hours to $10$ min using the hybrid approach. This approach should prove useful in the optimization of portions of complex or cascaded systems using gradient-based methods.

\bmsubsection*{Acknowledgments}
This work was funded by the French National Research Agency ANR Project META INCOME (ANR-24-CE91-0007).

\bmsubsection*{Conflicts of Interest}

The authors declare no conflicts of interest.

\bmsubsection*{Data Availability Statement}

The data, code and results used in this manuscript can be found in the Zenodo repository \cite{bensoltane_zenodo_2026}.


\begin{thebibliography}{10}

\bibitem{ma_Discrete_2005}
W.~Ma, M.R. Rayner, C.G. Parini, {\it IEEE Transactions on Antennas and
  Propagation} \textbf{2005}, {\it 53} (1), 339--346.

\bibitem{semouchkina_FDTD_2005}
E.A. Semouchkina, G.B. Semouchkin, M.~Lanagan, C.A. Randall, {\it IEEE
  Transactions on Microwave Theory and Techniques} \textbf{2005}, {\it 53} (4),
  1477--1487.

\bibitem{choi_Transmission_2011}
Wonjun Choi, Allard~P. Mosk, Q-Han Park, Wonshik Choi, {\it Physical Review B}
  \textbf{2011}, {\it 83} (13), 134207.

\bibitem{smith_Stochastic_2012}
Steven~M. Smith, Cynthia Furse, {\it IEEE Transactions on Antennas and
  Propagation} \textbf{2012}, {\it 60} (7), 3343--3350.

\bibitem{demesy_Scattering_2018}
Guillaume Dem{\'e}sy, Jean-Claude Auger, Brian Stout, {\it JOSA A}
  \textbf{2018}, {\it 35} (8), 1401--1409.

\bibitem{horodynski_Antireflection_2022a}
Michael Horodynski, Matthias K{\"u}hmayer, Cl{\'e}ment Ferise, Stefan Rotter,
  Matthieu Davy, {\it Nature} \textbf{2022}, {\it 607} (7918), 281--286.

\bibitem{niu_Stochastic_2024}
Kaikun Niu, Zijing Xiao, Guoda Xie, Xingang Ren, Yingsong Li, Zhixiang Huang,
  Xianliang Wu, Atef~Z. Elsherbeni, {\it IEEE Transactions on Microwave Theory
  and Techniques} \textbf{2024}, {\it 72} (7), 3935--3946.

\bibitem{haynes_Multipole_2011}
Mark Haynes, Mahta Moghaddam, {\it IEEE Transactions on Antennas and
  Propagation} \textbf{2011}, {\it 59} (1), 225--235.

\bibitem{grigoriev_Singular_2013}
V.~Grigoriev, S.~Varault, G.~Boudarham, B.~Stout, J.~Wenger, N.~Bonod, {\it
  Physical Review A} \textbf{2013}, {\it 88} (6), 063805.

\bibitem{pai_Scattering_2021a}
Pritam Pai, Jeroen Bosch, Matthias K{\"u}hmayer, Stefan Rotter, Allard~P. Mosk,
  {\it Nature Photonics} \textbf{2021}, {\it 15} (6), 431--434.

\bibitem{zhumabay_Inverse_2026}
Zhazira Zhumabay, Cl{\'e}ment Ferise, Vincent Pagneux, Stefan Rotter, Matthieu
  Davy, {\it Laser \& Photonics Reviews} \textbf{2026}, {\it 20} (11), e01055.

\bibitem{saillard_Scattering_1990}
M.~Saillard, D.~Maystre, {\it J. Opt. Soc. Am. A} \textbf{1990}, {\it 7} (6),
  982--990.

\bibitem{cangellarisNovelClosedformGreens2000}
A.C. Cangellaris, V.I. Okhmatovski, {\it IEEE Transactions on Microwave Theory
  and Techniques} \textbf{2000}, {\it 48} (12), 2225--2232.

\bibitem{chenNewGreensFunction2012}
Yongpin~P. Chen, Weng~Cho Chew, Lijun Jiang, {\it IEEE Transactions on Antennas
  and Propagation} \textbf{2012}, {\it 60} (10), 4766--4776.

\bibitem{zhangGeneralGreensFunction2017}
Shu-Hui Zhang, Wen Yang, Kai Chang, {\it Physical Review B} \textbf{2017}, {\it
  95} (7), 075421.

\bibitem{panGreensFunctionsGeophysics2019}
Ernian Pan, {\it Reports on Progress in Physics} \textbf{2019}, {\it 82} (10),
  106801.

\bibitem{boumrarEquivalenceWaveFunction2020}
Hocine Boumrar, Mahdi Hamidi, Hand Zenia, Samir Lounis, {\it Journal of
  Physics: Condensed Matter} \textbf{2020}, {\it 32} (35), 355302.

\bibitem{mosk_Controlling_2012}
Allard~P. Mosk, Ad~Lagendijk, Geoffroy Lerosey, Mathias Fink, {\it Nature
  Photonics} \textbf{2012}, {\it 6} (5), 283--292.

\bibitem{carminati_Speckle_2015}
R.~Carminati, G.~Cwilich, L.~S. {Froufe-P{\'e}rez}, J.~J. S{\'a}enz, {\it
  Physical Review A} \textbf{2015}, {\it 91} (2), 023807.

\bibitem{rotter_Light_2017}
Stefan Rotter, Sylvain Gigan, {\it Reviews of Modern Physics} \textbf{2017},
  {\it 89} (1), 015005.

\bibitem{delhougne_Coherent_2021}
Philipp del Hougne, K.~Brahima Yeo, Philippe Besnier, Matthieu Davy, {\it
  Physical Review Letters} \textbf{2021}, {\it 126} (19), 193903.

\bibitem{gigan_Roadmap_2022}
Sylvain Gigan, Ori Katz, Hilton~B {de Aguiar}, Esben~Ravn Andresen, Alexandre
  Aubry, Jacopo Bertolotti, Emmanuel Bossy, Dorian Bouchet, Joshua Brake,
  Sophie Brasselet, Yaron Bromberg, Hui Cao, Thomas Chaigne, Zhongtao Cheng,
  Wonshik Choi, Tom{\'a}{\v s} {\v C}i{\v z}m{\'a}r, Meng Cui, Vincent~R
  Curtis, Hugo Defienne, Matthias Hofer, Ryoichi Horisaki, Roarke Horstmeyer,
  Na~Ji, Aaron~K LaViolette, Jerome Mertz, Christophe Moser, Allard~P Mosk,
  Nicolas~C P{\'e}gard, Rafael Piestun, Sebastien Popoff, David~B Phillips,
  Demetri Psaltis, Babak Rahmani, Herv{\'e} Rigneault, Stefan Rotter, Lei Tian,
  Ivo~M Vellekoop, Laura Waller, Lihong Wang, Timothy Weber, Sheng Xiao, Chris
  Xu, Alexey Yamilov, Changhuei Yang, Hasan Y{\i}lmaz, {\it Journal of Physics:
  Photonics} \textbf{2022}, {\it 4} (4), 042501.

\bibitem{vynck_Light_2023}
Kevin Vynck, Romain Pierrat, R{\'e}mi Carminati, Luis~S. {Froufe-P{\'e}rez},
  Frank Scheffold, Riccardo Sapienza, Silvia Vignolini, Juan~Jos{\'e}
  S{\'a}enz, {\it Reviews of Modern Physics} \textbf{2023}, {\it 95} (4),
  045003.

\bibitem{liutkus_Imaging_2014}
Antoine Liutkus, David Martina, S{\'e}bastien Popoff, Gilles Chardon, Ori Katz,
  Geoffroy Lerosey, Sylvain Gigan, Laurent Daudet, Igor Carron, {\it Scientific
  Reports} \textbf{2014}, {\it 4} (1), 5552.

\bibitem{bertolotti_Imaging_2022}
Jacopo Bertolotti, Ori Katz, {\it Nature Physics} \textbf{2022}, {\it 18} (9),
  1008--1017.

\bibitem{delhougne_Experimental_2021}
Philipp del Hougne, R{\'e}mi Sobry, Olivier Legrand, Fabrice Mortessagne,
  Ulrich Kuhl, Matthieu Davy, {\it Laser \& Photonics Reviews} \textbf{2021},
  {\it 15} (3), 2000335.

\bibitem{davy_Mean_2021}
Matthieu Davy, Matthias K{\"u}hmayer, Sylvain Gigan, Stefan Rotter, {\it
  Communications Physics} \textbf{2021}, {\it 4} (1), 85.

\bibitem{dong_MetamaterialBased_2012}
Yuandan Dong, Tatsuo Itoh, {\it Proceedings of the IEEE} \textbf{2012}, {\it
  100} (7), 2271--2285.

\bibitem{jackson_LeakyWave_2012}
David~R. Jackson, Christophe Caloz, Tatsuo Itoh, {\it Proceedings of the IEEE}
  \textbf{2012}, {\it 100} (7), 2194--2206.

\bibitem{milias_MetamaterialInspired_2021}
Christos Milias, Rasmus~B. Andersen, Pavlos~I. Lazaridis, Zaharias~D. Zaharis,
  Bilal Muhammad, Jes T.~B. Kristensen, Albena Mihovska, Dan D.~S. Hermansen,
  {\it IEEE Access} \textbf{2021}, {\it 9}, 89846--89865.

\bibitem{zheng_LeakyWave_2023}
Dongze Zheng, Chi~Hou Chan, Ke~Wu, {\it IEEE Journal of Microwaves}
  \textbf{2023}, {\it 3} (1), 368--397.

\bibitem{lalau-keraly_Adjoint_2013}
Christopher~M. {Lalau-Keraly}, Samarth Bhargava, Owen~D. Miller, Eli
  Yablonovitch, {\it Optics Express} \textbf{2013}, {\it 21} (18),
  21693--21701.

\bibitem{hughes_Adjoint_2018a}
Tyler~W. Hughes, Momchil Minkov, Ian A.~D. Williamson, Shanhui Fan, {\it ACS
  Photonics} \textbf{2018}, {\it 5} (12), 4781--4787.

\bibitem{elbechari_Adjoint_2022}
Reda El~Bechari, Fr{\'e}d{\'e}ric Guyomarch, St{\'e}phane Brisset, {\it
  Mathematics} \textbf{2022}, {\it 10} (6), 885.

\bibitem{guillamon_Insitu_2025a}
John Guillamon, Cheng-Zhen Wang, Zin Lin, Tsampikos Kottos, {\it Nature
  Communications} \textbf{2025}, {\it 16} (1), 11466.

\bibitem{neviere_Electromagnetic_1996}
Michel Neviere, Frederic Montiel, \bblin{} {\it Multilayer and Grazing
  Incidence X-Ray/EUV Optics III}, (\capitalize\bbleds{}Richard~B. Hoover,
  Arthur B. C.~Walker II ), International Society for Optics and Photonics,
  SPIE, \textbf{1996}, \bblpp{}. 176 -- 183.

\bibitem{stout_Transfer_2002}
Brian Stout, Jean-Claude Auger, Jacques Lafait, {\it Journal of Modern Optics}
  \textbf{2002}, {\it 49} (13), 2129--2152.

\bibitem{hugonin_Hybridization_2008}
J.~P. Hugonin, M.~Besbes, P.~Lalanne, {\it Optics Letters} \textbf{2008}, {\it
  33} (14), 1590--1592.

\bibitem{asadova_Tmatrix_2025}
Nigar Asadova, Karim Achouri, Kristian Arjas, Baptiste Augui{\'e}, Roland
  Aydin, Alexandre Baron, Dominik Beutel, Bernd Bodermann, Kaoutar Boussaoud,
  Sven Burger, Minseok Choi, Krzysztof~M. Czajkowski, Andrey~B. Evlyukhin,
  Atefeh {Fazel-Najafabadi}, Ivan {Fernandez-Corbaton}, Puneet Garg, David
  Globosits, Ulrich Hohenester, Hongyoon Kim, Seokwoo Kim, Philippe Lalanne,
  Eric~C. Le~Ru, J{\"o}rg Meyer, Jungho Mun, Lorenzo Pattelli, Lukas Pflug,
  Carsten Rockstuhl, Junsuk Rho, Stefan Rotter, Brian Stout, P{\"a}ivi
  T{\"o}rm{\"a}, Jorge~Olmos Trigo, Frank Tristram, Nikolaos~L. Tsitsas, Renaud
  Vall{\'e}e, Kevin Vynck, Thomas Weiss, Peter Wiecha, Thomas Wriedt, Vassilios
  Yannopapas, Maxim~A. Yurkin, Grigorios~P. Zouros, {\it Journal of
  Quantitative Spectroscopy and Radiative Transfer} \textbf{2025}, {\it 333},
  109310.

\bibitem{chaumet_Modelling_2017}
Patrick~C. Chaumet, Anne Sentenac, Anne-Laure Fehrembach, {\it Optical and
  Quantum Electronics} \textbf{2017}, {\it 49} (2), 71.

\bibitem{markel_Extinction_2019}
Vadim~A. Markel, {\it Journal of Quantitative Spectroscopy and Radiative
  Transfer} \textbf{2019}, {\it 236}, 106611.

\bibitem{chaumetDiscreteDipoleApproximation2022}
Patrick~C. Chaumet, {\it Mathematics} \textbf{2022}, {\it 10} (17), 3049.

\bibitem{lippmann_Variational_1950}
B.~A. Lippmann, Julian Schwinger, {\it Physical Review} \textbf{1950}, {\it 79}
  (3), 469--480.

\bibitem{thouless_Conductivity_1981}
D.~J. Thouless, S.~Kirkpatrick, {\it Journal of Physics C: Solid State Physics}
  \textbf{1981}, {\it 14} (3), 235.

\bibitem{drouvelis_Parallel_2006}
P.~S. Drouvelis, P.~Schmelcher, P.~Bastian, {\it Journal of Computational
  Physics} \textbf{2006}, {\it 215} (2), 741--756.

\bibitem{davy_Selectively_2018}
Matthieu Davy, Azriel~Z. Genack, {\it Nature Communications} \textbf{2018},
  {\it 9} (1), 4714.

\bibitem{rivero_GreensFunctionDefect_2021}
Jose D.~H. Rivero, Li~Ge, {\it Physical Review B} \textbf{2021}, {\it 103}
  (19), 195142.

\bibitem{plancherel_Contribution_1910}
Michel Plancherel, Mittag Leffler, {\it Rendiconti del Circolo Matematico di
  Palermo (1884-1940)} \textbf{1910}, {\it 30} (1), 289--335.

\bibitem{bensoltane_zenodo_2026}
Isam BEN~SOLTANE, Romain ASSO, Matthieu Davy, {\it Fast local field calculation
  in cascaded coupled systems - Code and data}, \textbf{2026}.
  \urlprefix\url{https://doi.org/10.5281/zenodo.22097076}.

\end{thebibliography}

\providecommand{\url}[1]{\texttt{#1}}
\providecommand{\urlprefix}{}
\providecommand{\foreignlanguage}[2]{#2}
\providecommand{\Capitalize}[1]{\uppercase{#1}}
\providecommand{\capitalize}[1]{\expandafter\Capitalize#1}
\providecommand{\bibliographycite}[1]{\cite{#1}}
\providecommand{\bbland}{and}
\providecommand{\bblchap}{chap.}
\providecommand{\bblchapter}{chapter}
\providecommand{\bbletal}{et~al.}
\providecommand{\bbleditors}{editors}
\providecommand{\bbleds}{eds: }
\providecommand{\bbleditor}{editor}
\providecommand{\bbled}{ed.}
\providecommand{\bbledition}{edition}
\providecommand{\bbledn}{ed.}
\providecommand{\bbleidp}{page}
\providecommand{\bbleidpp}{pages}
\providecommand{\bblerratum}{erratum}
\providecommand{\bblin}{in}
\providecommand{\bblmthesis}{Master's thesis}
\providecommand{\bblno}{no.}
\providecommand{\bblnumber}{number}
\providecommand{\bblof}{of}
\providecommand{\bblpage}{page}
\providecommand{\bblpages}{pages}
\providecommand{\bblp}{p}
\providecommand{\bblphdthesis}{Ph.D. thesis}
\providecommand{\bblpp}{pp}
\providecommand{\bbltechrep}{}
\providecommand{\bbltechreport}{Technical Report}
\providecommand{\bblvolume}{volume}
\providecommand{\bblvol}{Vol.}
\providecommand{\bbljan}{January}
\providecommand{\bblfeb}{February}
\providecommand{\bblmar}{March}
\providecommand{\bblapr}{April}
\providecommand{\bblmay}{May}
\providecommand{\bbljun}{June}
\providecommand{\bbljul}{July}
\providecommand{\bblaug}{August}
\providecommand{\bblsep}{September}
\providecommand{\bbloct}{October}
\providecommand{\bblnov}{November}
\providecommand{\bbldec}{December}
\providecommand{\bblfirst}{First}
\providecommand{\bblfirsto}{1st}
\providecommand{\bblsecond}{Second}
\providecommand{\bblsecondo}{2nd}
\providecommand{\bblthird}{Third}
\providecommand{\bblthirdo}{3rd}
\providecommand{\bblfourth}{Fourth}
\providecommand{\bblfourtho}{4th}
\providecommand{\bblfifth}{Fifth}
\providecommand{\bblfiftho}{5th}
\providecommand{\bblst}{st}
\providecommand{\bblnd}{nd}
\providecommand{\bblrd}{rd}
\providecommand{\bblth}{th}

\bmsubsection*{Supporting Information}

Additional supporting information can be found online in the Supporting Information
section.


\end{document}